\documentclass[aps,prd,11pt,reprint,nofootinbib,showkeys]{revtex4-1}

\usepackage{setspace}
\usepackage{subfigure}
\usepackage{amsmath,amssymb,amsfonts,amsthm,mathrsfs}
\usepackage{graphicx,wrapfig}
\usepackage{enumerate} 

\newcommand{\be}{\nopagebreak[3]\begin{equation}}
\newcommand{\ee}{\end{equation}}
\newcommand{\bea}{\nopagebreak[3]\begin{eqnarray}}
\newcommand{\ea}{\end{eqnarray}}

\begin{document}

\title{Breaking of isospectrality of quasinormal modes in nonrotating loop quantum gravity black holes}

\author{Daniel del-Corral }
 \email{corral.martinez@ubi.pt}
\affiliation{Departamento de Física, Universidade da Beira Interior, Rua Marquês D'Ávila e Bolama, 6201-001 Covilhã, Portugal.}
\affiliation{Centro de Matemática e Aplicações da Universidade da Beira Interior, Rua Marquês D'Ávila e Bolama, 6201-001 Covilhã, Portugal.}%
\author{Javier Olmedo}
 \email{javolmedo@ugr.es}
\affiliation{Departamento de F\'isica Te\'orica y del Cosmos, Universidad de Granada, Granada-18071, Spain}%

\begin{abstract}
We study the quasinormal frequencies of three effective geometries of nonrotating regular black holes derived from loop quantum gravity. Concretely, we consider the Ashtekar-Olmedo-Singh and two Gambini-Olmedo-Pullin prescriptions. We compute the quasinormal frequencies of axial and polar perturbations adopting a WKB method. We show that they differ from those of classical general relativity and, more importantly, that isospectrality is broken. Nevertheless, these deviations are tiny, even for microscopic black holes, and they decay following an inverse power law of the size of the mass of the black holes. For the sake of completeness, we also analyze scalar and vector perturbations, reaching similar conclusions.

\end{abstract}
\maketitle
\section{Introduction}\label{sec:intro}

Quasinormal modes are one of the most interesting physical aspects of black hole space-times. They describe the behavior of these compact objects when they are perturbed, for instance, during the last stages of the ringdown regime after merging. These perturbations produce ripples in the fabric of space-time that propagate with a complex frequency outwards to infinity and inwards to the horizon of the black hole. The real part of these frequencies corresponds to time oscillations, while the imaginary one produces an exponential dissipation of the perturbations in time. Furthermore, in classical general relativity, these complex frequencies only depend on the mass, charge, and angular momentum of the black hole. In this way, they provide a way to test the validity of this classical description, as well as the consistency of modified theories of gravity and models motivated by quantum gravity theories. The direct detection of these frequencies is expected to be possible with sufficient precision in current gravitational wave detectors, such as LIGO and Virgo \cite{ligo}.

Perturbation theory of (nonrotating) spherically symmetric vacuum black holes was initially discussed by T. Regge and J.A. Wheeler \cite{reggewheeler}, for axial perturbations. Later on, Zerilli \cite{zerilli} provided a similar formulation of polar perturbations. In both cases, they correspondingly derived a second order differential equation for the radial part, which can be understood as a Schr\"odinger-like equation with a potential. Quasinormal modes were then discussed by Chandrasekhar and Detweiler \cite{qnm}. But more interestingly, they proved the isospectrality of axial and polar perturbations, namely, that they share the same quasinormal frequencies, despite each perturbation obeys a different radial equation. Actually, isospectrality has been also shown in Reissner-Nordstr\"om metric, and recently, this question was clarified for Schwarzschild-(anti)-de-Sitter geometries in \cite{iso-ads}. Moreover, recent work states that isospectrality of perturbations of Schwarzschild black holes has actually its origin in the Darboux covariance of the infinite set of possible master equations \cite{iso-db-cov}.

Interestingly, in some modified theories of gravity, isospectrality is violated (see for instance \cite{iso-vio1,iso-vio2,iso-vio3}), and general analyses show that isospectrality is actually very fragile \cite{iso-gen1,iso-gen2}. It has also been shown to be the case, for instance, in some effective geometries within loop quantum gravity \cite{polar-lqc} (see \cite{ashtekarsingh} for a review of this quantization program). Recently, other analysis \cite{lqc-qnm1,lqc-qnm2,lqc-qnm3} that calculated quasinormal frequencies of scalar, vector and (axial) tensor perturbations for effective geometries in loop quantum gravity did not discuss if isospectrality remains unbroken. This is an interesting question given that recent works on the instability of quasinormal modes \cite{instab1,instab2} show that the quasinormal mode spectrum of black holes is unstable under quite generic perturbations. On the other hand, a number of authors have suggested that quasinormal modes play a central role in the understanding of the black hole area spectrum \cite{magg,acrmp,carpi}. All this indicates that quasinormal mode physics might be a window to probe observationally fundamental high-frequency (even Planck scale) physics in the near future.

In this paper we compute the fundamental and some few overtones of quasinormal frequencies of scalar, vector and tensor perturbations (axial and polar), of three effective geometries of regular black holes in loop quantum gravity. In particular, we will focus on the Ashtekar-Olmedo-Singh (AOS) effective geometries \cite{aos,ao} and, on the other hand, on the Gambini-Olmedo-Pullin (GOP) prescriptions suggested in \cite{gop-imp1,gop-imp2}. We show that isospectrality of axial and polar quasinormal modes of these prescriptions is violated. However, this violation is tiny, as well as deviations from general relativity, even for very small black hole masses (a few thousands of the Planck mass), very close to the limiting regime of validity of those effective models.\footnote{A more realistic description of Planckian black holes in loop quantum gravity requires additional ingredients from the full theory \cite{lqg-bh,lqg-bh2,lqg-bh3}.} Actually, we see that quantum corrections to the quasinormal frequencies decay universally with a given power of the mass  of the black hole. 

The paper is organized as follows. In Sec. \ref{sec:effect} we introduce the effective geometries that will be studied in this manuscript. Sec. \ref{sec:qnm} contains the main findings about the computation of quasinormal frequencies (fundamental mode and a few overtones) of axial and polar perturbations, and discusses if isospectrality is broken. We conclude in Sec. \ref{sec:concl}. For the sake of completeness, we added Appendix \ref{app:qnm} where we compute a few quasinormal frequencies of scalar and vector perturbations; Appendix \ref{sec:pade} summarizes the implementation of Pad\'e approximants in the WKB method in order to improve its accuracy; and Appendix \ref{sec:tables} contains tables of quasinormal frequencies of scalar, vector and tensor (axial and polar) perturbations, as well as an error estimation of the WKB method. 

\section{Spherically symmetric effective geometries}\label{sec:effect}

We will focus on static spherically symmetric (effective) geometries described by the line element 
\begin{equation}\label{eq:ds2}
\mathrm{d} s^{2}=-G(r) \mathrm{d} t^{2}+F(r) \mathrm{d} r^{2}+H(r) \mathrm{d} \Omega^{2},
\end{equation}
with $\mathrm{d} \Omega^{2}=\mathrm{d} \theta^{2}+\sin^2 \theta \mathrm{d} \varphi^{2}$ the standard line element of the unit round sphere. Among several effective geometries within loop quantum gravity, we will focus here on three recent proposals:

\begin{itemize}
    \item Ashtekar-Olmedo-Singh (AOS): The effective geometries proposed in \cite{aos} (see \cite{ao} for additional details) are characterized by a space-time line element as in \eqref{eq:ds2} where
\begin{widetext}
    \begin{align}\nonumber
        &G_{AOS}(r) = \left(\frac{r}{r_{S}}\right)^{2 \epsilon} \frac{\left(1-\left(\frac{r_{S}}{r}\right)^{1+\epsilon}\right)\left(2+\epsilon+\epsilon\left(\frac{r_{S}}{r}\right)^{1+\epsilon}\right)^{2}\left((2+\epsilon)^{2}-\epsilon^{2}\left(\frac{r_{S}}{r}\right)^{1+\epsilon}\right)}{16\left(1+\frac{\delta_{c}^{2} L_{0}^{2} \gamma^{2} r_{S}^{2}}{16 r^{4}}\right)(1+\epsilon)^{4}},\\\nonumber
        &F_{AOS}(r) = \left(1+\frac{\delta_{c}^{2} L_{0}^{2} \gamma^{2} r_{S}^{2}}{16 r^{4}}\right) \frac{\left(\epsilon+\left(\frac{r}{r_{S}}\right)^{1+\epsilon}(2+\epsilon)\right)^{2}}{\left(\left(\frac{r}{r_{S}}\right)^{1+\epsilon}-1\right)\left(\left(\frac{r}{r_{S}}\right)^{1+\epsilon}(2+\epsilon)^{2}-\epsilon^{2}\right)},\\
        &H_{AOS}(r) = r^{2}\left(1+\frac{\gamma^{2} L_{0}^{2} \delta_{c}^{2} r_{S}^{2}}{16 r^{4}}\right),\label{eq:aos}
    \end{align}
with $r\in[r_S,\infty)$, $r_S=2Gm$ is the Schwarzschild radius in the radial coordinate $r$,
\begin{equation}
    L_{o} \delta_{c}=\frac{1}{2}\left(\frac{\gamma \Delta^{2}}{4 \pi^{2} m}\right)^{1 / 3},\quad \epsilon+1=\sqrt{1+\gamma^2\left(\frac{\sqrt{\Delta}}{\sqrt{2 \pi} \gamma^{2} m}\right)^{2 / 3}},
\end{equation}
\end{widetext}
are quantum parameters, with $L_0$ playing the role of an infrared regulator, $\gamma=0.2375$ is the Immirzi parameter, and $\Delta=5.17\ell_{\rm Pl}^2$ is the area gap (minimum nonzero eigenvalue of the area operator in loop quantum gravity). This effective geometry was shown to be valid for macroscopic black holes, namely, at least black holes with masses a few orders of magnitude larger than the Planck mass $M_{\rm Pl}=\sqrt{\hbar/G}$ or larger. 

\item Gambini-Olmedo-Pullin (GOP): These effective geometries were given in \cite{gop-imp1,gop-imp2} (see \cite{gp,gop,bh-rev} for additional details). Their space-time line element of the form \eqref{eq:ds2} is characterized by
\begin{align}\nonumber
        &G^{\alpha}_{GOP}(r) = \left(1-\frac{r_{S}}{r+r_{0}}+\alpha\frac{\tilde\Delta}{4 \pi} \frac{r_{S}^{4}}{\left(r+r_{0}\right)^{6}\left(1+\frac{r_{S}}{r+r_{0}}\right)^{2}}\right),\\\nonumber
        &F^{\alpha}_{GOP}(r) = \frac{\left(1+\frac{\delta x}{2\left(r+r_{0}\right)}\right)^{2}}{\left(1-\frac{r_{S}}{r+r_{0}}+\alpha\frac{\tilde\Delta}{4 \pi} \frac{r_{S}^{4}}{\left(r+r_{0}\right)^{6}\left(1+\frac{r_{S}}{r+r_{0}}\right)^{2}}\right)},\\
        &H_{GOP}(r) = (r+r_0)^{2},\label{eq:gop}
    \end{align}
with $r\in[r_H,\infty)$ where $r_H$ represents the radial position of the black hole horizon on these effective geometries (for instance, if $\alpha=0$ then $r_H=r_S-r_0$ with $r_S=2Gm$),  and 
\begin{equation}
    r_0 = \left(\frac{2 G m \tilde \Delta}{4 \pi}\right)^{1 / 3},
\end{equation}
where $\tilde\Delta=21.77\ell_{Pl}^2$ is the area gap in loop quantum gravity with unit Immirzi parameter ($\gamma=1$) and $\delta x\in[\ell_{\rm Pl}^2/(2r_0),r_0]$ is a quantum parameter that codifies the discreteness of the radial coordinate. Here, we choose $\delta x=r_0$, as oppose to the choice $\delta x = \ell^2_{\rm Pl}/r_0$ that would give the smoothest effective geometry (we will also discuss this choice in Sec. \ref{sec:qnm}). Besides, $\alpha$ is a parameter that takes the values 0 or 1. If $\alpha=1$, the resulting effective geometry corresponds to the one initially proposed in \cite{gop-imp1} (see also Ref. \cite{kswe-imp} for a similar proposal). However, in \cite{gop-imp2}, the choice $\alpha=0$ was suggested in order to alleviate some undesirable slicing dependence emerging in these effective geometries.\footnote{Ref. \cite{lqc-qnm3} takes into account $\alpha=1$ for the computation of quasinormal frequencies.} Here, we will consider both proposals, since the slicing dependent corrections in the prescription $\alpha=1$ are negligible outside the horizon. 

\end{itemize}

\section{quasinormal modes}\label{sec:qnm}

Besides the original studies of Refs. \cite{reggewheeler,zerilli},  gauge invariant perturbations in spherically symmetric space-times have been studied under the canonical framework \cite{moncrief} (see also \cite{briz1,briz2} for the nonvacuum case), and adopting a 2+2 decomposition \cite{gersen,sar-tiglio,marpoiss,2p2}.\footnote{Despite \cite{2p2} adopts the Regge-Wheeler gauge in their study, the master equations derived there can be cast into a gauge invariant formulation (see also \cite{marpoiss}).} They satisfy second order differential hyperbolic equations. Their time-dependence, since the background is static, can be factorized out. Similarly, the background is spherically symmetric. Hence, perturbations can be expanded in a basis of scalar, vector and tensor spherical harmonics. Within this decomposition, tensor modes can be classified in two groups according to the tensor harmonics polarity: there are axial (also denoted as odd, magnetic or toroidal) harmonics and polar (or also known as even, electric or poloidal) harmonics. At the end of the day, one is left with the radial part of the second order differential equation, which takes the form 
\begin{equation}
    \frac{\partial^2\psi_{\tilde\omega,\ell}}{\partial r_{*}^{2}}+\left[\tilde\omega^2-V_\ell(r)\right] \psi_{\tilde\omega,\ell}=0, \label{diffeq}
\end{equation}
with $\tilde\omega$ the (dimensionful) 
frequency and $V_\ell(r)$ an effective potential. Besides, $r_{*}$ is the tortoise coordinate defined as\footnote{Its definition is such that the 1+1 part of the space-time metric, after ignoring the angular one, is conformal to a 1+1 Minkowski metric. }
\begin{equation}
     \mathrm{d} r_* = \sqrt{\frac{F(r)}{G(r)}} \mathrm{d} r.
\end{equation}
For instance, in the Schwarzschild classical space-time, it takes the form 
\begin{equation}
r_* = r + r_S\ln\left(\frac{r}{r_S} - 1\right), \end{equation}
with $r\in (r_S,\infty)$. 

Quasinormal modes are solutions to Eq. \eqref{diffeq} with boundary conditions
\begin{align}\nonumber
\psi_{n,\ell}(r)&\propto e^{-i \tilde\omega_{n,\ell} r_*}\quad\quad r_*\to+\infty, \\
\psi_{n,\ell}(r)&\propto e^{i \tilde\omega_{n,\ell} r_*}\quad\quad \;\;r_*\to-\infty,\label{eq:qnm-bndry}
\end{align}
which actually require that frequencies $\tilde\omega_{n,\ell}$ belong to a discrete subset of   imaginary numbers, with the imaginary part of $\tilde\omega_{n,\ell}$ being positive. In order to compute these quasinormal frequencies, we will adopt a WKB method. It allows one to find approximate solutions to Eq. \eqref{diffeq}. Here, one studies the solutions in three regions: the two covering the intervals where the potential is negligible, namely, at spatial infinity and close to the horizon, which amounts to $r_*\to\pm\infty$, and the central one, where the effective potential reaches its peak. In the asymptotic regions, the solutions to Eq. \eqref{diffeq} are approximated by a linear combination of exponential functions which are matched with the solutions in the central region. Here, the potential is approximated by a Taylor expansion. The corresponding solutions to this approximated equation can be computed approximately. Once the matching is performed, one relates the ingoing and outgoing amplitudes, obtaining a pair of connection formulas. Then, applying the boundary conditions \eqref{eq:qnm-bndry} one finally finds the following formula (see \cite{wkb} for details) that allows us to compute the complex quasinormal frequencies
\begin{equation}
\tilde\omega_{n,\ell}^2=V_{\ell}(\tilde r)-\sqrt{-2V_{\ell}^{''}(\tilde r)}\left[\left(n+\frac12\right)+\sum_{i=2}^N \Lambda^{(i)}_{n,\ell}(\tilde r)\right]. \label{wkbformula}
\end{equation}
The functions $\Lambda^{(i)}_{n,\ell}(\tilde r)$ codify high-order WKB corrections, which depend exclusively on the derivatives of the potential evaluated at the peak $\tilde r$ (the effective potentials considered here have a single peak located at $\tilde r\simeq 3/2 r_S$).\footnote{The concrete value of $\tilde r$ not only depends on the type of perturbation but also on the multipole $\ell$, and for the effective geometries AOS and GOP, there is also an extra dependence on the mass of the black hole. } They have a rather complicated expression but explicit formulas are given in \cite{wkb} up to third order and in \cite{konoplya} up to sixth order. In addition to these high-order approximations, we consider Pad\'e approximants in order to increase the accuracy of our computations (see Appendix \ref{sec:pade} for details). We found that a 10th WKB order combined with Pad\'e approximants gives enough accuracy for our purposes. All the quasinormal frequencies and error estimations have been computed using the \textit{Mathematica} code given in \cite{code}. Moreover, we report our numerical results in terms of dimensionless frequencies defined as $\omega_{n,\ell}=(r_S/2G)\, \tilde \omega_{n,\ell}$. 

\subsection{Axial perturbations}\label{sec:axial}

The potential of axial perturbations can be expressed in terms of the metric components of the line element in Eq. \eqref{eq:ds2} as
\begin{equation}
 {}^{(a)}V_\ell(r)=G(r)\left[\frac{\ell(\ell+1)}{H(r)}-R(r)\right],
\end{equation}
where
\begin{align}\label{eq:Rfunc}\nonumber
    R(r)=&\frac{2}{H(r)}+\frac{1}{F(r)}\bigg(\frac{G'(r)H'(r)}{4G(r)H(r)}-\frac{F'(r)H'(r)}{4F(r)H(r)}\\
&-\frac{3[H'(r)]^2}{4H^2(r)}+\frac{H''(r)}{2H(r)}\bigg).
\end{align}
The primes denote derivative with respect to $r$. This potential agrees with expressions given, for instance, in \cite{briz1,wang,lqc-qnm1} (however it differs from the one adopted by \cite{lqc-qnm2}). For the Schwarzschild geometry, that we denote in the following by GR, it agrees with the standard result
\begin{equation}
    {}^{(a)}V^{GR}_\ell(r) = \left(1-\frac{2G m}{r}\right)\left[\frac{\ell(\ell+1)}{r^{2}}-\frac{6 G m}{r^{3}}\right].
\end{equation}
Here, we focus on generic static spherically symmetric space-times. We ignore contributions from the (possibly effective) matter sector that could modify the potential ${}^{(a)}V_\ell(r)$ and/or induce couplings with polar perturbations. We expect that those contributions will be negligible with respect to the ones already incorporated in our effective description.

\begin{figure}[ht]
{\centering     
  \includegraphics[width = 0.49\textwidth]{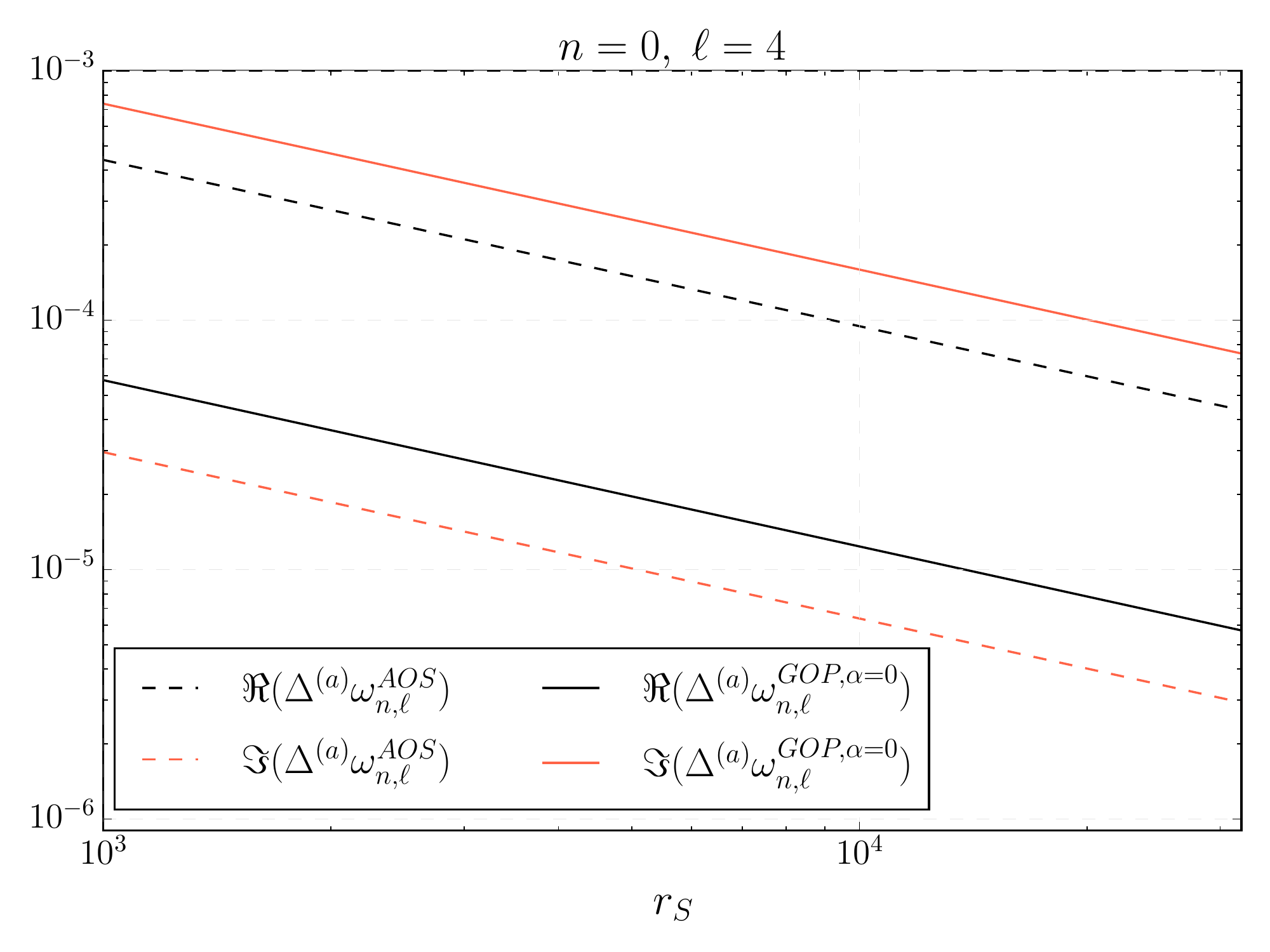}
}
 \caption{In this plot we show the deviation of the real (black) and imaginary (red) parts of the quasinormal frequency $n=0$ and $\ell=4$ of AOS (dashed line) and GOP with $\alpha=0$ (solid line) effective geometries with respect to general relativity. For this frequency our numerical methods give good enough accuracy. We do not show the GOP prescription with $\alpha=1$ since it gives qualitatively similar results than GOP one with $\alpha=0$. }
\label{fig:lqg-vs-gr}
\end{figure}

In Table \ref{axial} we show the dimensionless quasinormal frequencies for the axial perturbations of GR, AOS and GOP metrics. We denote them by the symbol ${}^{(a)}\omega_{n,\ell}$. In both AOS and GOP metrics we take a black hole with Schwarzschild radius $r_S=10^3\,\ell_P$. In the AOS geometry this corresponds to a value of the characteristic parameter of $\epsilon\cong2.848\times 10^{-3}$. For GOP we consider the two prescriptions ($\alpha=0$ and $\alpha=1$). Both show a similar spectrum of quasinormal frequencies. These values of the horizon radius lay in the limiting regime where we expect that these effective geometries still provide a reliable description. In all cases, our findings have enough accuracy to show deviations from GR. Depending on the multipole $\ell$ and $n$, this deviation barely exceeds one part in one thousand in the best case, even for these tiny black holes. We also observe that these deviations decrease with the mass of the black hole. For instance, in Fig. \ref{fig:lqg-vs-gr} we show, for axial perturbations, the absolute differences $\Delta{}^{(a)}\omega_{n,\ell}^{AOS}=({}^{(a)}\omega^{GR}_{n,\ell}-{}^{(a)}\omega^{AOS}_{n,\ell})$  and $\Delta {}^{(a)}\omega_{n,\ell}^{GOP,\alpha=0}=({}^{(a)}\omega^{GR}_{n,\ell}-{}^{(a)}\omega^{GOP,\alpha=0}_{n,\ell})$, of a given overtone, as functions of the horizon radius of the black holes. Here we consider black holes with $r_S\in[1000 \ell_{\rm Pl},32000\ell_{\rm Pl}]$. We observe that deviations from general relativity actually decrease with the power $(r_S/\ell_{\rm Pl})^{-2/3}$. Therefore, they are negligible for macroscopic (i.e. solar mass) black holes. It is worth mentioning that for the GOP prescriptions this behavior depends on the choice of the quantum parameter $\delta x$. We have also analyzed the choice $\delta x=\ell_{\rm Pl}^2/(2r_0)$, which amounts to smoother geometries and hence smaller quantum corrections. Although here we do not show the results, we checked that in this case quantum corrections in the quasinormal frequencies decrease as $(r_S/\ell_{\rm Pl})^{-4/3}$. For the fundamental mode and other overtones we have observed the same behaviour discussed above.

\subsection{Polar perturbations}\label{sec:polar}

For polar perturbations, the potential has a more complicated expression than the axial ones. It takes the form
\begin{equation}
\begin{split}
    {}^{(p)}V_\ell(r)&=\frac{G(r) (\ell-1)^2(\ell+2)^2}{\lambda_\ell(r)^2}\left[\frac{(\ell-1)(\ell+2)+2}{H(r)}+R(r)\right.\\&+\left.\frac{H(r)R(r)^2}{(\ell-1)^2(\ell+2)^2}\left((\ell-1)(\ell+2)+\frac{H(r)R(r)}{3}\right)\right],
\end{split}
\end{equation}
where $R(r)$ is again given by \eqref{eq:Rfunc} and $\lambda_\ell(r)$ by
\begin{equation}
    \lambda_\ell(r)=(\ell-1)(\ell+2)+H(r)R(r).
\end{equation}
One can see that the expression above agrees with the one obtained by \cite{2p2} (but it disagrees with the potential of polar perturbations derived in \cite{polar-lqc}). By direct inspection, one can also check that it reduces to the standard one in the case of the Schwarzschild classical geometry given by
\begin{equation}
\begin{split}
    {}^{(p)}V^{GR}_\ell(r) &= \left(1-\frac{2 Gm}{r}\right)\bigg[\frac{\ell(\ell+1)}{r^{2}}\\
&-\frac{6 Gm}{r^{3}} \frac{r^{2} \Lambda_\ell(\Lambda_\ell+2)+3 Gm(r-Gm)}{(r \Lambda_\ell+3 Gm)^{2}}\bigg],
\end{split}
\end{equation}
with $\Lambda_\ell=(\ell-1)(\ell+2)/2$. In Table \ref{polar} we provide a few dimensionless quasinormal frequencies for polar perturbations of GR, AOS and GOP metrics. We denote them by ${}^{(p)}\omega_{n,\ell}$. We adopt the same choice of parameters than for quasinormal frequencies of axial perturbations, namely, a Schwarzschild radius $r_S=10^3\,\ell_P$ and the two GOP prescriptions ($\alpha=0$ and $\alpha=1$). Again, we see quantitative deviations with respect to the Schwarzschild geometry, but they barely exceed one part in one thousand. Their behavior is very similar to the one of axial perturbations shown in Fig. \ref{fig:lqg-vs-gr}. Up to a small percent, their deviation with respect to general relativity is of the same order of magnitude and it decays as $(r_S/\ell_{\rm Pl})^{-2/3}$. Again, for the GOP prescriptions, if instead we choose $\delta x=\ell_{\rm Pl}^2/(2r_0)$, quantum corrections decrease as $(r_S/\ell_{\rm Pl})^{-4/3}$. We also checked that several overtones and the fundamental mode share all those properties.

\begin{figure}[b]
{\centering     
  \includegraphics[width = 0.49\textwidth]{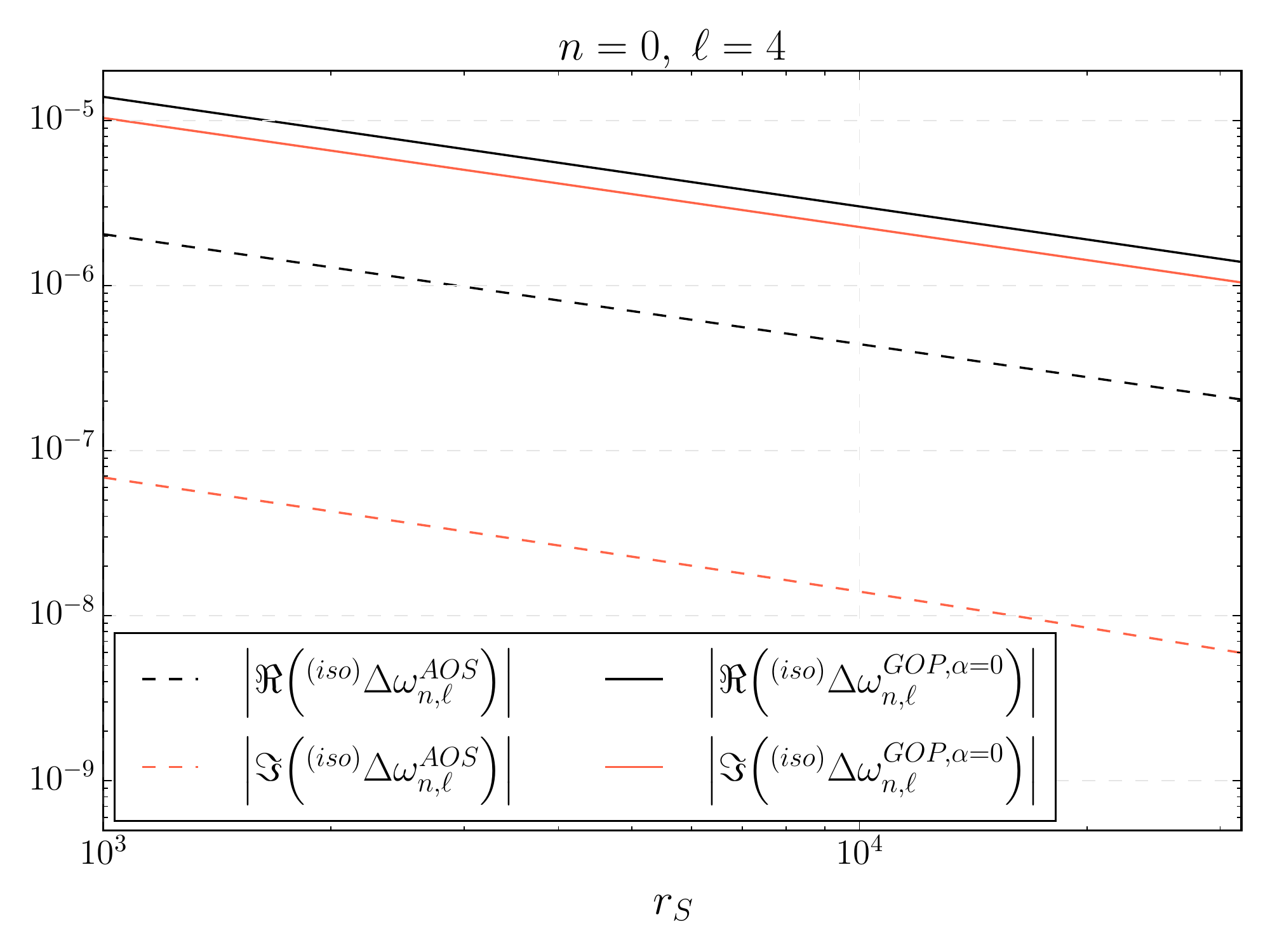}
}
 \caption{In this plot we show the deviation from isospectrality of the (absolute value of the) real (black) and imaginary (red) parts of the quasinormal frequency $n=0$ and $\ell=4$ of AOS (dashed line) and GOP with $\alpha=0$ (solid line) effective geometries. Our numerical methods give good enough accuracy for this frequency. The GOP prescription with $\alpha=1$ gives qualitatively similar results than $\alpha=0$. Hence, we do not show it here.}
\label{fig:lqg-iso}
\end{figure}

Interestingly, our results also allow us to probe the isospectrality of axial and polar quasinormal modes of these effective geometries. Within our accuracy, we see isospectrality of quasinormal frequencies in the Schwarzschild classical metric. This is not surprising, since it is well known that quasinormal frequencies of Schwarzschild classical geometries are isospectral. In our findings, for these classical geometries, any differences between axial and polar modes are always some orders of magnitude below our error estimation. However, this is not the case of the AOS and GOP regular black holes. Here, isospectrality is broken.\footnote{We estimate the error of our calculations in Table \ref{errors}. These errors are qualitatively smaller than the isospectrality violation of these effective geometries.} Although the violation is very weak, even for these microscopic black holes, the lowest multipole $n=0$ and $\ell=2$ shows the strongest deviation from isospectrality, which happens well above our numerical errors. In Fig. \ref{fig:lqg-iso}, as an example, we show the deviation from isospectrality codified in ${}^{(iso)}\Delta \omega_{n,\ell}^{AOS}=\left({}^{(a)}\omega^{AOS}_{n,\ell}-{}^{(p)}\omega^{AOS}_{n,\ell}\right)$, for a given overtone, as a function of the horizon radius of the black holes, and similarly for the GOP ($\alpha=0$) prescription ---the GOP one for $\alpha=1$ shows qualitatively similar results. We consider the radius to be in the interval $r_S\in[1000\,\ell_{\rm Pl},32000\,\ell_{\rm Pl}]$. We observe that the violation of isospectrality also decreases with the power $(r_S/\ell_{\rm Pl})^{-2/3}$. This behavior seems to be universal regardless of the choice of $n$ and $\ell$ within the modes that we have been able to probe. Hence, for macroscopic black holes, isospectrality is expected to be preserved except for very tiny deviations. In the case of the GOP prescriptions, these corrections decrease as $(r_S/\ell_{\rm Pl})^{-4/3}$ if one chooses instead $\delta x=\ell_{\rm Pl}^2/(2r_0)$.

\section{Conclusions}\label{sec:concl}

In this manuscript we study the quasinormal frequencies of axial and polar perturbations of several black hole effective geometries in loop quantum gravity. Concretely, we consider the AOS prescription proposed in \cite{aos}, as well as the GOP prescriptions derived in \cite{gop-imp1,gop-imp2}. The effective spacetime line elements are characterized by the functions given in \eqref{eq:aos} and \eqref{eq:gop}, respectively. We compare them with the Schwarzschild classical space-time. We provide expressions for the effective potential of the radial equations of axial and polar modes valid for general spherically symmetric geometries. Moreover, we complement our study with a similar calculation of scalar and vector perturbations in Appendix \ref{app:qnm}.

In order to compute the quasinormal frequencies, we adopt a high-order WKB approach supplemented with Pad\'e approximants that allow us to increase considerably the accuracy of our calculations. We compare  classical and quantum corrected black holes with the same black hole horizon radius. On the one hand, the difference of the quasinormal frequencies between the classical and effective geometries decays always for AOS and GOP prescriptions as $(r_S/\ell_{\rm Pl})^{-2/3}$ if, for the latter, one chooses the quantum parameter $\delta x=r_0$ in the GOP prescriptions. However, for the GOP one, these quantum corrections  decay as $(r_S/\ell_{\rm Pl})^{-4/3}$ if one chooses $\delta x=\ell_{\rm Pl}^2/(2r_0)$. Besides, we have also demonstrated numerically that these effective geometries break isopectrality. Namely, quasinormal frequencies of axial and polar perturbations disagree, unlike in the classical theory. However, this violation also dilutes with the radius (i.e. ADM mass) of the black hole as $(r_S/\ell_{\rm Pl})^{-2/3}$ (for AOS and GOP prescriptions if $\delta x=r_0$), or as $(r_S/\ell_{\rm Pl})^{-4/3}$ (for the GOP prescriptions if $\delta x=\ell_{\rm Pl}^2/(2r_0)$). Hence, one should expect large deviations only for microscopic black holes. Nevertheless for macroscopic black holes, our study shows that they are negligible. 

It is worth mentioning that we adopt an effective approach that assumes a continuous background geometry and also a particular choice for the effective potentials of the radial equations of axial and polar perturbations that ignores quantum fluctuations of the geometries and couplings originated from other quantum corrections. We must note that GOP effective geometries were derived from a quantum theory, for a particular family of semiclassical states. For instance, they neglect superpositions in the mass and the basis of spin networks. These superpositions will add additional corrections on these effective geometries that are worth to be explored in the future. Besides, these potentials actually suffer from regularization and quantization ambiguities. Despite all that, we expect that our results are robust regarding the breaking of isospectrality, which is the main result of this manuscript. Besides, although we have considered just a few effective geometries, we expect that isospectrality will also be broken in other black hole models \cite{kswe-imp,achou,gssw,bmm,abv,qmm}.  

\acknowledgements

We acknowledge D. Brizuela, R. Gambini, J. L. Jaramillo, J. Martos, J. Pullin and C. Sopuerta for very useful comments. This work is supported by the Spanish Government through the projects PID2020-118159GB-C43, PID2019-105943GB-I00 (with FEDER contribution), and the "Operative Program FEDER2014-2020 Junta de Andaluc\'ia-Consejer\'ia de Econom\'ia y Conocimiento" under project E-FQM-262-UGR18 by Universidad de Granada. Daniel del-Corral is grateful for the support of grant
UI/BD/151491/2021 from the Portuguese Agency Fundação para a Ciência e a Tecnologia.
This research was funded by Fundação para a Ciência e a Tecnologia grant number
UIDB/MAT/00212/2020.

\appendix

\section{quasinormal modes of scalar and vector perturbations}\label{app:qnm}

In this Appendix we also compute the quasinormal frequencies of scalar and vector (electromagnetic) perturbations. We follow the same methodology adopted for the axial and polar modes. Expressions for the corresponding potentials can be derived straightforwardly (see for instance \cite{aagls}).

The effective potential for scalar (massless) perturbations can be obtained in terms of the metric components as:
\begin{equation}
{}^{(s)}V_\ell(r)=G(r)\left[\frac{\ell(\ell+1)-2}{H(r)}+\frac{[H'(r)]^2}{2F(r)H^2(r)}+R(r)\right],
\end{equation}
where we must recall that the primes denote derivatives with respect to $r$. For the classical Schwarzschild solution, one obtains
\begin{equation}
    {}^{(s)}V^{GR}_\ell(r) = \left(1-\frac{2G m}{r}\right)\left[\frac{\ell(\ell+1)}{r^{2}}+\frac{2 G m}{r^{3}}\right].
\end{equation}

For vector (massless) perturbations the potential takes a simple form:
\begin{equation}
    {}^{(v)}V_\ell(r)=G(r)\,\frac{\ell(\ell+1)}{H(r)}.
\end{equation}
This expression reduces to 
\begin{equation}
    {}^{(v)}V^{GR}_\ell(r) = \left(1-\frac{2G m}{r}\right)\frac{\ell(\ell+1)}{r^{2}},
\end{equation}
for the classical Schwarzschild geometry.

\begin{figure}[ht]
{\centering     
  \includegraphics[width = 0.49\textwidth]{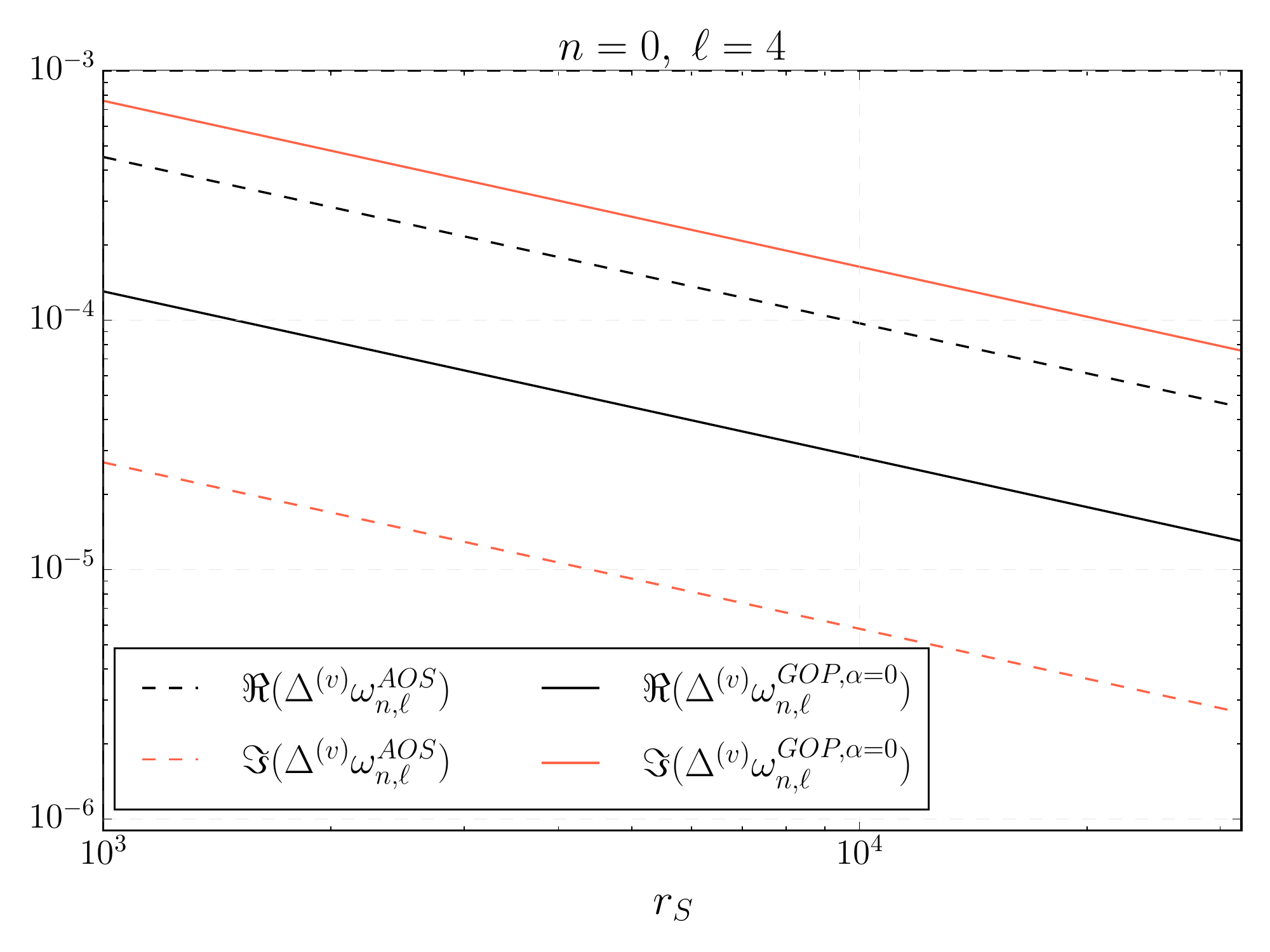}
  \includegraphics[width = 0.49\textwidth]{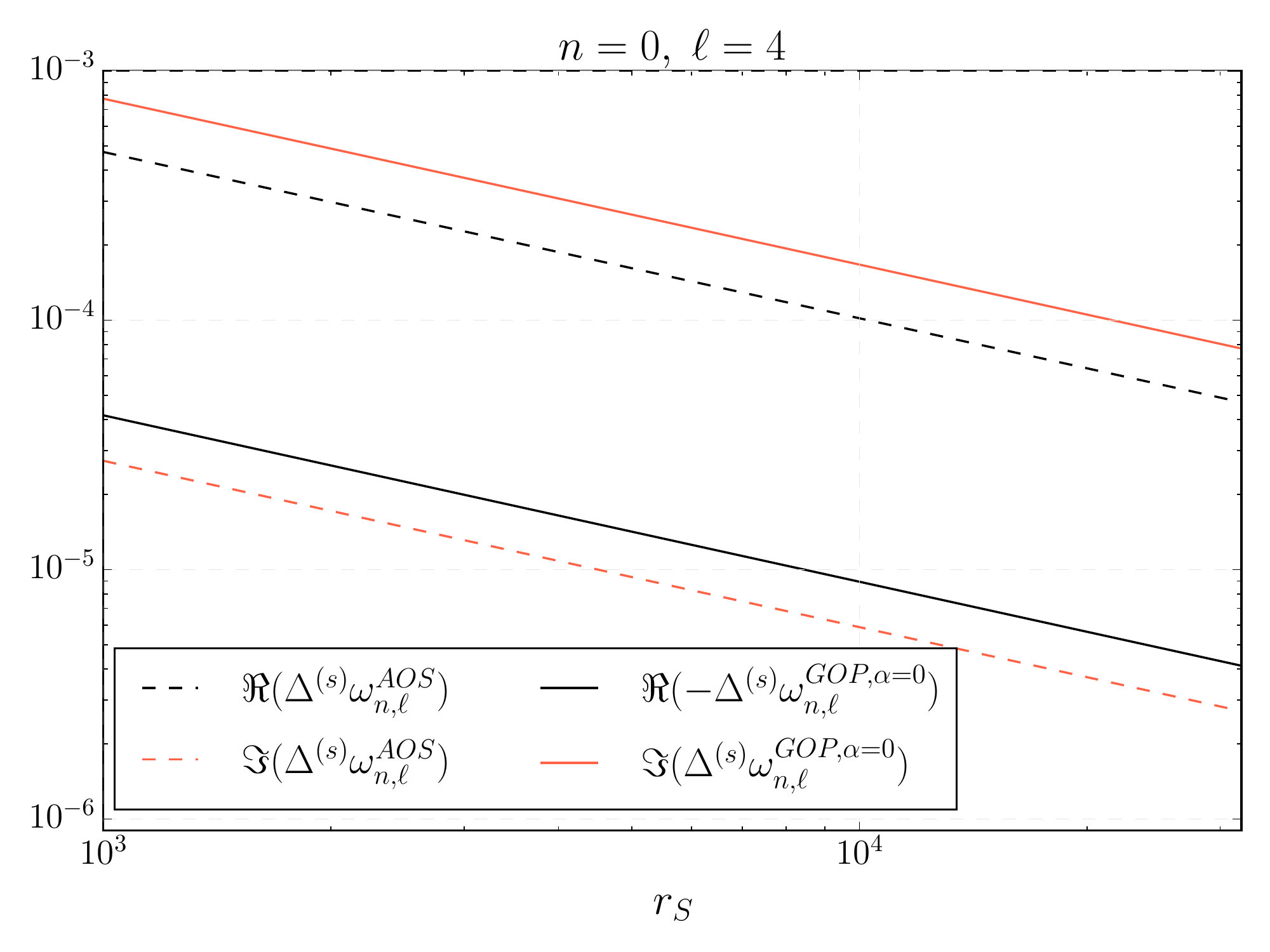}
}
 \caption{In this plot we show the deviation of the real (black) and imaginary (red) parts of the quasinormal frequency $n=0$ and $\ell=4$ of AOS (dashed line) and GOP with $\alpha=0$ (solid line) effective geometries with respect to general relativity. In the upper panel we show this difference for vector perturbations. In the lower panel for scalar perturbations. In both cases, our computations give good enough accuracy for this frequency. We do not show the GOP prescription with $\alpha=1$ since it gives qualitatively similar results than GOP one with $\alpha=0$. }
\label{fig:lqg-vs-gr-vs}
\end{figure}

In Table \ref{scalar} and Table \ref{vector} we show some quasinormal frequencies for the scalar ${}^{(s)}\omega_{n,\ell}$ and vector ${}^{(v)}\omega_{n,\ell}$ perturbations of GR, AOS and GOP metrics. Again, we have considered black holes with horizon radius equal or close to $r_S=10^3\,\ell_P$. As in the axial and polar cases, the deviations from GR are small, but still the accuracy of our method is good enough to show this corrections. In Fig. \ref{fig:lqg-vs-gr-vs} we see that the differences $\Delta{}^{(v)}\omega_{n,\ell}^{AOS}=({}^{(v)}\omega^{GR}_{n,\ell}-{}^{(v)}\omega^{AOS}_{n,\ell})$  and $\Delta {}^{(v)}\omega_{n,\ell}^{GOP,\alpha=0}=({}^{(v)}\omega^{GR}_{n,\ell}-{}^{(v)}\omega^{GOP,\alpha=0}_{n,\ell})$, for vector perturbations, and $\Delta{}^{(s)}\omega_{n,\ell}^{AOS}=({}^{(s)}\omega^{GR}_{n,\ell}-{}^{(s)}\omega^{AOS}_{n,\ell})$  and $\Delta {}^{(s)}\omega_{n,\ell}^{GOP,\alpha=0}=({}^{(s)}\omega^{GR}_{n,\ell}-{}^{(s)}\omega^{GOP,\alpha=0}_{n,\ell})$, for scalar perturbations, weaken with the mass of the black holes as $(r_S/\ell_{\rm Pl})^{-2/3}$. 

\section{Pad\'e approximants and accuracy of the WKB method}\label{sec:pade}

In order to increase the accuracy of the WKB method it is possible to make use of the Pad\'e approximants. Let us summarize the procedure (for details see \cite{wkbpade,konoplyapade}). Here, one starts with the polynomial 
\begin{equation}
    P_k(\xi)=V_{\ell}(r_0)-\sqrt{-2V_{\ell}^{''}(r_0)}\left[\xi\left(n+\frac12\right)+\sum_{n=2}^N \xi^n\Lambda_n\right],
\end{equation}
from the formula \eqref{wkbformula} where the parameter $\xi$ is introduced to allow us to keep track of the order of the polynomial, so that $\tilde\omega_{n,\ell}^2=P_k(1)$, where $k$ is the WKB order. Then, we approximate $P_k(\xi)$ around $\xi=0$ with the family of the Pad\'e approximants 
\begin{equation}
    P_{\tilde{n}/\tilde{m}}(\xi)=\frac{Q_0+Q_1\,\xi+...+Q_{\tilde{n}}\,\xi^{\tilde{n}}}{R_0+R_1\,\xi+...+R_{\tilde{m}}\,\xi^{\tilde{m}}},
\end{equation}
where $\tilde{n}+\tilde{m}=k$ and $Q_i, R_i$ are determined such that 
\begin{equation}
    P_{\tilde{n}/\tilde{m}}(\xi)-P_k(\xi)=\mathcal{O}(\xi^{k+1}).
\end{equation}
Once an approximation $P_{\tilde{n}/\tilde{m}}(\xi)$ has been chosen, to compute the quasinormal frequencies, one takes $\tilde\omega_n^2=P_{\tilde{n}/\tilde{m}}(1)$. For example, at first order, the WKB formula admits two expressions
\begin{equation}
    \begin{split}
        \tilde\omega_n^2&=P_{1/0}(1)=P_1(1)=V_\ell(r_0)-\sqrt{-V''_\ell(r_0)}\left(n+\frac12\right),\\
        \tilde\omega_n^2&=P_{0/1}(1)=\frac{V_\ell(r_0)}{V_\ell(r_0)-\sqrt{-V_\ell''(r_0)}\left(n+\frac12\right)}.
    \end{split}
\end{equation}
For higher orders, it has been observed \cite{wkbpade,konoplyapade} that the Pad\'e approximants $P_{\tilde{n}/\tilde{m}}$ with $\tilde{n}\approx\tilde{m}$ usually give better results than the standard WKB formula $P_{\tilde{m}/0}$. Also, as the WKB order increases, all the alternative approximations give similar results. 

We use \cite{code} to compute the quasinormal frequencies together with an estimation of their error. The procedure is as follows. If we consider, for instance, the 10th WKB order, there are 11 Pad\'e approximants ($P_{0/10}$, $P_{1/9}$, ..., $P_{10/0}$). The criterion we adopt to select the most accurate approximants is the same as in \cite{konoplyapade}.  Here, one computes all the mean values of two closest frequencies 
\begin{equation}
    \bar \omega_{\tilde{n}/\tilde{m}}^{(p)}=\frac{\tilde\omega_{(\tilde{n}+p)/(\tilde{m}-p)}+\tilde\omega_{\tilde{n}/\tilde{m}}}{2},
\end{equation}
for $p$ a given set of integers that will be selected by the WKB approximation. For instance, for 
the 10th WKB order $p=1,\ldots,4$. Then for each value of $p$, we select the two estimations $\bar \omega_{\tilde{n}/\tilde{m}}^{(p)}$ for which the relative difference between the corresponding Pad\'e approximants (or frequencies) is minimum. Eventually, the final quasinormal frequency is the mean of these eight Pad\'e approximants and its absolute error given by the standard deviation of this set of estimations. We denote them by $\sigma_{axial}$ and $\sigma_{polar}$ for the axial and polar quasinormal frequencies, respectively. The code \cite{code} follows this criterion, providing mean values and errors (see \cite{konoplyapade} for details) as we just explained.

The estimations of the errors given in Table \ref{errors} are calculated in the following way: given $\sigma_{axial}$ and $\sigma_{polar}$, we define the total error as 
\begin{equation}
    \sigma=\sqrt{\sigma_{axial}^2+\sigma_{polar}^2}.
\end{equation}
We then compute the percentage relative error over the absolute value of the mean between the axial and polar quasinormal frequencies. 

\begin{widetext}
\section{Tables}\label{sec:tables}

In this appendix we provide numerical values of some of the dimensionless quasinormal frequencies, namely, we show $\omega_{n,\ell}=(r_S/2G)\,\tilde\omega_{n,\ell}$, as is customary in the literature. 

\begingroup
\setlength{\tabcolsep}{10pt}
\renewcommand{\arraystretch}{0.5}
\begin{table}[h]
\footnotesize
\centering
\begin{tabular}{|ccc|}
\hline\hline
\multicolumn{3}{|c|}{\textbf{Axial Perturbations}}                                                                                     \\ \hline\hline
\multicolumn{1}{|c|}{($n$,$\ell$)} & \multicolumn{1}{c|}{\textbf{Schwarzschild}}                     & \textbf{AOS} ($r_S=10^3\,\ell_P$)    \\ \hline
\multicolumn{1}{|c|}{(0,2)}      & \multicolumn{1}{c|}{0.74733225 - 0.17792806$i$}           & 0.74771876 - 0.17787563$i$             \\ \hline
\multicolumn{1}{|c|}{(1,2)}      & \multicolumn{1}{c|}{0.69322645 - 0.54811740$i$}           & 0.69360839 - 0.54794196$i$             \\ \hline
\multicolumn{1}{|c|}{(0,3)}      & \multicolumn{1}{c|}{1.19888658 - 0.18540612$i$}           & 1.19942465 - 0.18535672$i$            \\ \hline
\multicolumn{1}{|c|}{(1,3)}      & \multicolumn{1}{c|}{1.16528891 - 0.56259764$i$}           & 1.16581711 - 0.56244359$i$            \\ \hline
\multicolumn{1}{|c|}{(0,4)}      & \multicolumn{1}{c|}{1.61835676 - 0.18832792$i$}           & 1.61905472 - 0.18828092$i$           \\ \hline
\multicolumn{1}{|c|}{(1,4)}      & \multicolumn{1}{c|}{1.59326305 - 0.56866870$i$}           & 1.59395164 - 0.56852466$i$           \\ \hline
\multicolumn{1}{|c|}{(0,5)}      & \multicolumn{1}{c|}{2.02459062 - 0.18974103$i$}           & 2.02544902 - 0.18969535$i$           \\ \hline
\multicolumn{1}{|c|}{(1,5)}      & \multicolumn{1}{c|}{2.00444206 - 0.57163476$i$}           & 2.00529218 - 0.57149586$i$          \\ \hline
\multicolumn{1}{|c|}{(0,6)}      & \multicolumn{1}{c|}{2.42401964 - 0.19053169$i$}           & 2.42503820 - 0.19048680$i$         \\ \hline
\multicolumn{1}{|c|}{(1,6)}      & \multicolumn{1}{c|}{2.40714795 - 0.57329985$i$}           & 2.40815924 - 0.57316390$i$         \\ \hline
\multicolumn{1}{|c|}{}           & \multicolumn{1}{c|}{\textbf{GOP} ($\alpha=0$, $r_S=10^3\,\ell_P$)} & \textbf{GOP} ($\alpha=1$, $r_S=10^3\,\ell_P$) \\ \hline
\multicolumn{1}{|c|}{(0,2)}      & \multicolumn{1}{c|}{0.74736483 - 0.17720680$i$}           & 0.74736491 - 0.17720674$i$         \\ \hline
\multicolumn{1}{|c|}{(1,2)}      & \multicolumn{1}{c|}{0.69401982 - 0.54578773$i$}           & 0.69401995 - 0.54578751$i$         \\ \hline
\multicolumn{1}{|c|}{(0,3)}      & \multicolumn{1}{c|}{1.19894618 - 0.18468114$i$}           & 1.19894629 - 0.18468108$i$          \\ \hline
\multicolumn{1}{|c|}{(1,3)}      & \multicolumn{1}{c|}{1.16577313 - 0.56034440$i$}           & 1.16577326 - 0.56034419$i$           \\ \hline
\multicolumn{1}{|c|}{(0,4)}      & \multicolumn{1}{c|}{1.61841428 - 0.18758921$i$}           & 1.61841442 - 0.18758914$i$          \\ \hline
\multicolumn{1}{|c|}{(1,4)}      & \multicolumn{1}{c|}{1.59363598 - 0.56640627$i$}           & 1.59363614 - 0.56640607$i$          \\ \hline
\multicolumn{1}{|c|}{(0,5)}      & \multicolumn{1}{c|}{2.02464204 - 0.18899367$i$}           & 2.02464221 - 0.18899361$i$        \\ \hline
\multicolumn{1}{|c|}{(1,5)}      & \multicolumn{1}{c|}{2.00474707 - 0.56936206$i$}           & 2.00474725 - 0.56936186$i$          \\ \hline
\multicolumn{1}{|c|}{(0,6)}      & \multicolumn{1}{c|}{2.42406539 - 0.18977889$i$}           & 2.42406539 - 0.18977889$i$         \\ \hline
\multicolumn{1}{|c|}{(1,6)}      & \multicolumn{1}{c|}{2.40740646 - 0.57101967$i$}           & 2.40740646 - 0.57101967$i$          \\ \hline
\end{tabular}
\caption{Quasinormal frequencies for the first overtones of the axial perturbations.}
\label{axial}
\end{table}
\endgroup

\begingroup
\setlength{\tabcolsep}{10pt}
\renewcommand{\arraystretch}{0.5}
\begin{table}[h]
\footnotesize
\centering
\begin{tabular}{|ccc|}
\hline\hline
\multicolumn{3}{|c|}{\textbf{Polar Perturbations}}                                                                                     \\ \hline\hline
\multicolumn{1}{|c|}{($n$,$\ell$)} & \multicolumn{1}{c|}{\textbf{Schwarzschild}}                     & \textbf{AOS} ($r_S=10^3\,\ell_P$)    \\ \hline
\multicolumn{1}{|c|}{(0,2)}      & \multicolumn{1}{c|}{0.74734291 - 0.17792545$i$}           & 0.74770137 - 0.17786831$i$             \\ \hline
\multicolumn{1}{|c|}{(1,2)}      & \multicolumn{1}{c|}{0.69337238 - 0.54785794$i$}           & 0.69371455 - 0.54766125$i$             \\ \hline
\multicolumn{1}{|c|}{(0,3)}      & \multicolumn{1}{c|}{1.19888656 - 0.18540609$i$}           & 1.19941686 - 0.18535615$i$            \\ \hline
\multicolumn{1}{|c|}{(1,3)}      & \multicolumn{1}{c|}{1.16528740 - 0.56259608$i$}           & 1.16580600 - 0.56243957$i$            \\ \hline
\multicolumn{1}{|c|}{(0,4)}      & \multicolumn{1}{c|}{1.61835675 - 0.18832792$i$}           & 1.61905147 - 0.18828080$i$           \\ \hline
\multicolumn{1}{|c|}{(1,4)}      & \multicolumn{1}{c|}{1.59326306 - 0.56866870$i$}           & 1.59394788 - 0.56852416$i$           \\ \hline
\multicolumn{1}{|c|}{(0,5)}      & \multicolumn{1}{c|}{2.02459062 - 0.18974103$i$}           & 2.02544735 - 0.18969532$i$           \\ \hline
\multicolumn{1}{|c|}{(1,5)}      & \multicolumn{1}{c|}{2.00444206 - 0.57163476$i$}           & 2.00529032 - 0.57149572$i$          \\ \hline
\multicolumn{1}{|c|}{(0,6)}      & \multicolumn{1}{c|}{2.42401964 - 0.19053169$i$}           & 2.42503722 - 0.19048679$i$         \\ \hline
\multicolumn{1}{|c|}{(1,6)}      & \multicolumn{1}{c|}{2.40714795 - 0.57329985$i$}           & 2.40815818 - 0.57316385$i$         \\ \hline
\multicolumn{1}{|c|}{}           & \multicolumn{1}{c|}{\textbf{GOP} ($\alpha=0$, $r_S=10^3\,\ell_P$)} & \textbf{GOP} ($\alpha=1$, $r_S=10^3\,\ell_P$) \\ \hline
\multicolumn{1}{|c|}{(0,2)}      & \multicolumn{1}{c|}{0.74749081 - 0.17733983$i$}           & 0.74749089 - 0.17733977$i$         \\ \hline
\multicolumn{1}{|c|}{(1,2)}      & \multicolumn{1}{c|}{0.69407330 - 0.54597173$i$}           & 0.69407343 - 0.54597152$i$         \\ \hline
\multicolumn{1}{|c|}{(0,3)}      & \multicolumn{1}{c|}{1.19897954 - 0.18471131$i$}           & 1.19897964 - 0.18471125$i$          \\ \hline
\multicolumn{1}{|c|}{(1,3)}      & \multicolumn{1}{c|}{1.16577795 - 0.56043389$i$}           & 1.16577808 - 0.56043369$i$           \\ \hline
\multicolumn{1}{|c|}{(0,4)}      & \multicolumn{1}{c|}{1.61842820 - 0.18759958$i$}           & 1.61842834 - 0.18759952$i$          \\ \hline
\multicolumn{1}{|c|}{(1,4)}      & \multicolumn{1}{c|}{1.59364290 - 0.56643745$i$}           & 1.59364305 - 0.56643725$i$          \\ \hline
\multicolumn{1}{|c|}{(0,5)}      & \multicolumn{1}{c|}{2.02464917 - 0.18899817$i$}           & 2.02464934 - 0.18899811$i$        \\ \hline
\multicolumn{1}{|c|}{(1,5)}      & \multicolumn{1}{c|}{2.00475176 - 0.56937559$i$}           & 2.00475194 - 0.56937540$i$          \\ \hline
\multicolumn{1}{|c|}{(0,6)}      & \multicolumn{1}{c|}{2.42406933 - 0.18978122$i$}           & 2.42406953 - 0.18978115$i$         \\ \hline
\multicolumn{1}{|c|}{(1,6)}      & \multicolumn{1}{c|}{2.40740936 - 0.57102668$i$}           & 2.40740957 - 0.57102648$i$          \\ \hline
\end{tabular}
\caption{Quasinormal frequencies for the first overtones of the polar perturbations.}
\label{polar}
\end{table}
\endgroup

\begingroup
\setlength{\tabcolsep}{10pt}
\renewcommand{\arraystretch}{0.5}
\begin{table}[h]
\footnotesize
\centering
\begin{tabular}{|ccc|}
\hline\hline
\multicolumn{3}{|c|}{\textbf{Scalar Perturbations}}                                                                                     \\ \hline\hline
\multicolumn{1}{|c|}{($n$,$\ell$)} & \multicolumn{1}{c|}{\textbf{Schwarzschild}}                     & \textbf{AOS} ($r_S=10^3\,\ell_P$)    \\ \hline
\multicolumn{1}{|c|}{(0,0)}      & \multicolumn{1}{c|}{0.22106137 - 0.20937335$i$}           & 0.22140540 - 0.20925624$i$         \\ \hline
\multicolumn{1}{|c|}{(1,0)}      & \multicolumn{1}{c|}{0.18850053 - 0.70053554$i$}           & 0.18859010 - 0.70054267$i$             \\ \hline
\multicolumn{1}{|c|}{(0,1)}      & \multicolumn{1}{c|}{0.58587217 - 0.19532140$i$}           & 0.58621916 - 0.19527153$i$            \\ \hline
\multicolumn{1}{|c|}{(1,1)}      & \multicolumn{1}{c|}{0.52916747 - 0.61270000$i$}           & 0.52946242 - 0.61254038$i$            \\ \hline
\multicolumn{1}{|c|}{(0,2)}      & \multicolumn{1}{c|}{0.96728773 - 0.19351755$i$}           & 0.96775106 - 0.19347264$i$           \\ \hline
\multicolumn{1}{|c|}{(1,2)}      & \multicolumn{1}{c|}{0.92770066 - 0.59120829$i$}           & 0.92813871 - 0.59106793$i$           \\ \hline
\multicolumn{1}{|c|}{(0,3)}      & \multicolumn{1}{c|}{1.35073247 - 0.19299926$i$}           & 1.35133550 - 0.19295544$i$           \\ \hline
\multicolumn{1}{|c|}{(1,3)}      & \multicolumn{1}{c|}{1.32134299 - 0.58456958$i$}           & 1.32192959 - 0.58443479$i$          \\ \hline
\multicolumn{1}{|c|}{(0,4)}      & \multicolumn{1}{c|}{1.73483128 - 0.19278338$i$}           & 1.73558221 - 0.19273997$i$         \\ \hline
\multicolumn{1}{|c|}{(1,4)}      & \multicolumn{1}{c|}{1.71161607 - 0.58175204$i$}           & 1.71235477 - 0.58161967$i$         \\ \hline
\multicolumn{1}{|c|}{}           & \multicolumn{1}{c|}{\textbf{GOP} ($\alpha=0$, $r_S=10^3\,\ell_P$)} & \textbf{GOP} ($\alpha=1$, $r_S=10^3\,\ell_P$) \\ \hline
\multicolumn{1}{|c|}{(0,0)}      & \multicolumn{1}{c|}{0.22065868 - 0.20832678$i$}           & 0.22065871 - 0.20832671$i$         \\ \hline
\multicolumn{1}{|c|}{(1,0)}      & \multicolumn{1}{c|}{0.18841088 - 0.69652318$i$}           & 0.18841087 - 0.69652280$i$         \\ \hline
\multicolumn{1}{|c|}{(0,1)}      & \multicolumn{1}{c|}{0.58574403 - 0.19449923$i$}           & 0.58574409 - 0.19449915$i$          \\ \hline
\multicolumn{1}{|c|}{(1,1)}      & \multicolumn{1}{c|}{0.52965894 - 0.60989852$i$}           & 0.52965911 - 0.60989817$i$           \\ \hline
\multicolumn{1}{|c|}{(0,2)}      & \multicolumn{1}{c|}{0.96721207 - 0.19273001$i$}           & 0.96721215 - 0.19272994$i$          \\ \hline
\multicolumn{1}{|c|}{(1,2)}      & \multicolumn{1}{c|}{0.92812583 - 0.58870416$i$}           & 0.92812594 - 0.58870394$i$          \\ \hline
\multicolumn{1}{|c|}{(0,3)}      & \multicolumn{1}{c|}{1.35067881 - 0.19222161$i$}           & 1.35067892 - 0.19222154$i$        \\ \hline
\multicolumn{1}{|c|}{(1,3)}      & \multicolumn{1}{c|}{1.32166241 - 0.58216341$i$}           & 1.32166254 - 0.58216321$i$          \\ \hline
\multicolumn{1}{|c|}{(0,4)}      & \multicolumn{1}{c|}{1.73478970 - 0.19200982$i$}           & 1.73478985 - 0.19200976$i$         \\ \hline
\multicolumn{1}{|c|}{(1,4)}      & \multicolumn{1}{c|}{1.71186939 - 0.57938690$i$}           & 1.71186954 - 0.57938670$i$          \\ \hline
\end{tabular}
\caption{Quasinormal frequencies for the first overtones of the scalar perturbations.}
\label{scalar}
\end{table}
\endgroup

\begingroup
\setlength{\tabcolsep}{10pt}
\renewcommand{\arraystretch}{0.5}
\begin{table}[h]
\footnotesize
\centering
\begin{tabular}{|ccc|}
\hline\hline
\multicolumn{3}{|c|}{\textbf{Vector Perturbations}}                                                                                     \\ \hline\hline
\multicolumn{1}{|c|}{($n$,$\ell$)} & \multicolumn{1}{c|}{\textbf{Schwarzschild}}                     & \textbf{AOS} ($r_S=10^3\,\ell_P$)    \\ \hline
\multicolumn{1}{|c|}{(0,1)}      & \multicolumn{1}{c|}{0.49652620 - 0.18497687$i$}           & 0.49677857 - 0.18493659$i$             \\ \hline
\multicolumn{1}{|c|}{(1,1)}      & \multicolumn{1}{c|}{0.42896489 - 0.58735295$i$}           & 0.42917124 - 0.58722013$i$             \\ \hline
\multicolumn{1}{|c|}{(0,2)}      & \multicolumn{1}{c|}{0.91519104 - 0.19000886$i$}           & 0.91559456 - 0.18996650$i$            \\ \hline
\multicolumn{1}{|c|}{(1,2)}      & \multicolumn{1}{c|}{0.87308476 - 0.58142022$i$}           & 0.87346490 - 0.58128680$i$            \\ \hline
\multicolumn{1}{|c|}{(0,3)}      & \multicolumn{1}{c|}{1.31379734 - 0.19123244$i$}           & 1.31435699 - 0.19118977$i$           \\ \hline
\multicolumn{1}{|c|}{(1,3)}      & \multicolumn{1}{c|}{1.28347488 - 0.57945680$i$}           & 1.28401884 - 0.57932526$i$           \\ \hline
\multicolumn{1}{|c|}{(0,4)}      & \multicolumn{1}{c|}{1.70619039 - 0.19171987$i$}           & 1.70690733 - 0.19167710$i$           \\ \hline
\multicolumn{1}{|c|}{(1,4)}      & \multicolumn{1}{c|}{1.68253412 - 0.57862935$i$}           & 1.68323922 - 0.57849884$i$          \\ \hline
\multicolumn{1}{|c|}{(0,5)}      & \multicolumn{1}{c|}{2.09582556 - 0.19196334$i$}           & 2.09670029 - 0.19192054$i$         \\ \hline
\multicolumn{1}{|c|}{(1,5)}      & \multicolumn{1}{c|}{2.07644178 - 0.57820772$i$}           & 2.07730694 - 0.57807779$i$         \\ \hline
\multicolumn{1}{|c|}{}           & \multicolumn{1}{c|}{\textbf{GOP} ($\alpha=0$, $r_S=10^3\,\ell_P$)} & \textbf{GOP} ($\alpha=1$, $r_S=10^3\,\ell_P$) \\ \hline
\multicolumn{1}{|c|}{(0,1)}      & \multicolumn{1}{c|}{0.49512802 - 0.18174827$i$}           & 0.49512801 - 0.18174817$i$         \\ \hline
\multicolumn{1}{|c|}{(1,1)}      & \multicolumn{1}{c|}{0.42344855 - 0.57974580$i$}           & 0.42344839 - 0.57974555$i$         \\ \hline
\multicolumn{1}{|c|}{(0,2)}      & \multicolumn{1}{c|}{0.91441759 - 0.18693022$i$}           & 0.91441764 - 0.18693012$i$          \\ \hline
\multicolumn{1}{|c|}{(1,2)}      & \multicolumn{1}{c|}{0.86956543 - 0.57304523$i$}           & 0.86956536 - 0.57304497$i$           \\ \hline
\multicolumn{1}{|c|}{(0,3)}      & \multicolumn{1}{c|}{1.31325754 - 0.18818198$i$}           & 1.31325763 - 0.18818188$i$          \\ \hline
\multicolumn{1}{|c|}{(1,3)}      & \multicolumn{1}{c|}{1.28092560 - 0.57075050$i$}           & 1.28092561 - 0.57075023$i$          \\ \hline
\multicolumn{1}{|c|}{(0,4)}      & \multicolumn{1}{c|}{1.70577457 - 0.18867960$i$}           & 1.70577469 - 0.18867951$i$        \\ \hline
\multicolumn{1}{|c|}{(1,4)}      & \multicolumn{1}{c|}{1.68053934 - 0.56977960$i$}           & 1.68053940 - 0.56977932$i$          \\ \hline
\multicolumn{1}{|c|}{(0,5)}      & \multicolumn{1}{c|}{2.09548701 - 0.18892794$i$}           & 2.09548717 - 0.18892784$i$         \\ \hline
\multicolumn{1}{|c|}{(1,5)}      & \multicolumn{1}{c|}{2.07480446 - 0.56928366$i$}           & 2.07480456 - 0.56928338$i$          \\ \hline
\end{tabular}
\caption{Quasinormal frequencies for the first overtones of vector perturbations.}
\label{vector}
\end{table}
\endgroup

\begingroup
\setlength{\tabcolsep}{10pt}
\renewcommand{\arraystretch}{0.5}
\begin{table}[h]
\footnotesize
\centering
\begin{tabular}{|ccc|}
\hline\hline
\multicolumn{3}{|c|}{\textbf{Estimation of the errors (\%)}}                                                                                     \\ \hline\hline
\multicolumn{1}{|c|}{($n$,$\ell$)} & \multicolumn{1}{c|}{\textbf{Schwarzschild}}                     & \textbf{AOS} ($r_S=10^3\,\ell_P$)    \\ \hline
\multicolumn{1}{|c|}{(0,2)}      & \multicolumn{1}{c|}{8.32073$\times10^{-4}$}           & 8.28774$\times10^{-4}$             \\ \hline
\multicolumn{1}{|c|}{(1,2)}      & \multicolumn{1}{c|}{5.37259$\times10^{-2}$}           & 5.34523$\times10^{-2}$             \\ \hline
\multicolumn{1}{|c|}{(0,3)}      & \multicolumn{1}{c|}{3.78222$\times10^{-6}$}           & 3.65464$\times10^{-6}$            \\ \hline
\multicolumn{1}{|c|}{(1,3)}      & \multicolumn{1}{c|}{2.43954$\times10^{-5}$}           & 2.47351$\times10^{-5}$            \\ \hline
\multicolumn{1}{|c|}{(0,4)}      & \multicolumn{1}{c|}{4.13316$\times10^{-8}$}           & 4.02860$\times10^{-8}$           \\ \hline
\multicolumn{1}{|c|}{(1,4)}      & \multicolumn{1}{c|}{1.06444$\times10^{-6}$}           & 1.02453$\times10^{-6}$           \\ \hline
\multicolumn{1}{|c|}{(0,5)}      & \multicolumn{1}{c|}{1.83706$\times10^{-9}$}           & 1.65610$\times10^{-9}$           \\ \hline
\multicolumn{1}{|c|}{(1,5)}      & \multicolumn{1}{c|}{6.01152$\times10^{-8}$}           & 5.93357$\times10^{-8}$          \\ \hline
\multicolumn{1}{|c|}{(0,6)}      & \multicolumn{1}{c|}{5.17149$\times10^{-11}$}           & 4.51964$\times10^{-11}$        \\ \hline
\multicolumn{1}{|c|}{(1,6)}      & \multicolumn{1}{c|}{8.12115$\times10^{-9}$}           & 8.06923$\times10^{-9}$         \\ \hline
\multicolumn{1}{|c|}{}           & \multicolumn{1}{c|}{\textbf{GOP} ($\alpha=0$, $r_S=10^3\,\ell_P$)} & \textbf{GOP} ($\alpha=1$, $r_S=10^3\,\ell_P$) \\ \hline
\multicolumn{1}{|c|}{(0,2)}      & \multicolumn{1}{c|}{9.04563$\times10^{-4}$}           & 9.04640$\times10^{-4}$        \\ \hline
\multicolumn{1}{|c|}{(1,2)}      & \multicolumn{1}{c|}{5.08751$\times10^{-2}$}           & 5.08751$\times10^{-2}$         \\ \hline
\multicolumn{1}{|c|}{(0,3)}      & \multicolumn{1}{c|}{2.56594$\times10^{-6}$}           & 2.56612$\times10^{-6}$          \\ \hline
\multicolumn{1}{|c|}{(1,3)}      & \multicolumn{1}{c|}{2.35593$\times10^{-5}$}           & 2.35524$\times10^{-5}$           \\ \hline
\multicolumn{1}{|c|}{(0,4)}      & \multicolumn{1}{c|}{2.26845$\times10^{-8}$}           & 2.26960$\times10^{-8}$  \\ \hline
\multicolumn{1}{|c|}{(1,4)}      & \multicolumn{1}{c|}{6.71517$\times10^{-7}$}           & 6.70473$\times10^{-7}$          \\ \hline
\multicolumn{1}{|c|}{(0,5)}      & \multicolumn{1}{c|}{5.11426$\times10^{-10}$}           & 5.09071$\times10^{-10}$        \\ \hline
\multicolumn{1}{|c|}{(1,5)}      & \multicolumn{1}{c|}{5.05602$\times10^{-8}$}           & 5.04985$\times10^{-8}$          \\ \hline
\multicolumn{1}{|c|}{(0,6)}      & \multicolumn{1}{c|}{2.28438$\times10^{-11}$}           & 2.31124$\times10^{-11}$         \\ \hline
\multicolumn{1}{|c|}{(1,6)}      & \multicolumn{1}{c|}{7.27388$\times10^{-9}$}           & 7.26642$\times10^{-9}$          \\ \hline
\end{tabular}
\caption{Estimations of the errors using the 10th WKB order with Pad\'e approximants.}
\label{errors}
\end{table}
\endgroup

\clearpage

\end{widetext}


\end{document}